\documentclass[useAMS,usenatbib]{mn2e_arxiv_submit}

\usepackage{graphicx}
\usepackage{natbib}
\usepackage{graphicx}
\usepackage{color} 
\usepackage{pdfpages}
\usepackage{appendix}
\usepackage{subfigure}
\usepackage{amsmath}
\usepackage{amssymb}

\begin{document}
\bibliographystyle{aabib}
\renewcommand{\labelitemi}{$\bullet$}
\def\py{\textsc{Python}}
\def\tar{\textsc{Tardis} }
\def\cld{\textsc{Cloudy} }
\def\civ{C~\textsc{iv} }
\def\araa{ARAA}
\def\nat{Nature}
\def\apjl{ApJ Letters}
\def\aapr{AAPR}
\def\actaa{ACTAA} 
\def\ssr{SSR}
\def\apj{ApJ}
\def\apss{AP\&SS}
\def\pasp{PASP}
\def\aap{A\&A}
\def\mnras{MNRAS}
\def\aj{AJ}
\def\rmxaa{RMXAA}

\def\heiiopt{He~\textsc{ii}~$\lambda4686$}
\def\heiiuv{He~\textsc{ii}~$\lambda1640$}
\def\heiioptnew{He~\textsc{ii}~$\lambda3202$}
\def\la{Ly$\alpha$}
\def\ha{H$\alpha$}
\def\hb{H$\beta$}
\def\civfull{C~\textsc{iv}~$\lambda1550$}

%
%

\title{The Impact of Accretion Disk Winds on the Optical Spectra of Cataclysmic Variables}

\author[Matthews et al.]{J.~H.~Matthews,$^1$\thanks{E-mail: jm8g08@soton.ac.uk} C.~Knigge,$^1$ K.~S.~Long,$^2$ S.~A.~Sim,$^3$ and N.~Higginbottom$^4$
\medskip  
\\$^1$School of Physics and Astronomy, University of Southampton, Highfield, Southampton, SO17 1BJ, UK
\\$^2$Space Telescope Science Institute, 3700 San Martin Drive, Baltimore, MD 21218, USA
\\$^3$School of Mathematics and Physics, Queens University Belfast, University Road, Belfast, BT7 1NN, Northern Ireland, UK
\\$^4$Department of Astronomy, University of Las Vegas, Las Vegas, NV 89119, USA}

\date{Accepted, 15 April 2015. Received, 8 April 2015; in original form, 26 January 2015.}

%
%

\maketitle
\begin{abstract}
Many high-state non-magnetic cataclysmic variables (CVs) exhibit
blue-shifted absorption or P-Cygni profiles associated with ultraviolet
(UV) resonance lines. These features imply the existence of 
powerful accretion disk winds in CVs. Here, we use our Monte Carlo ionization
and radiative transfer code to investigate whether disk wind models that
produce realistic UV line profiles are also likely to generate
observationally significant recombination line and continuum emission
in the {\em optical} waveband. We also test whether outflows may be responsible for 
the single-peaked emission line profiles often seen in high-state CVs and for the weakness
of the Balmer absorption edge (relative to simple models of optically thick accretion disks).
We find that a standard disk wind model that is successful in
reproducing the UV spectra of CVs also leaves a noticeable imprint on
the optical spectrum, particularly for systems viewed at high
inclination. The strongest optical wind-formed recombination lines are \ha\ and
\heiiopt. We demonstrate that a higher-density outflow model produces 
all the expected H and He lines and produces a recombination continuum that can 
fill in the Balmer jump at high inclinations.
This model displays reasonable verisimilitude with the optical spectrum of RW Trianguli. No single-peaked emission is seen, although we observe a 
narrowing of the double-peaked emission lines from the base of the wind.
Finally, we show that even denser models can produce a single-peaked \ha\ line.
On the basis of our results, we suggest that winds can 
modify, and perhaps even dominate, the line and continuum emission
from CVs.
\end{abstract}


%
%

\section{Introduction} 
\label{sec:intro}

\begin{figure*}
\centering
\includegraphics[width=0.7\textwidth]{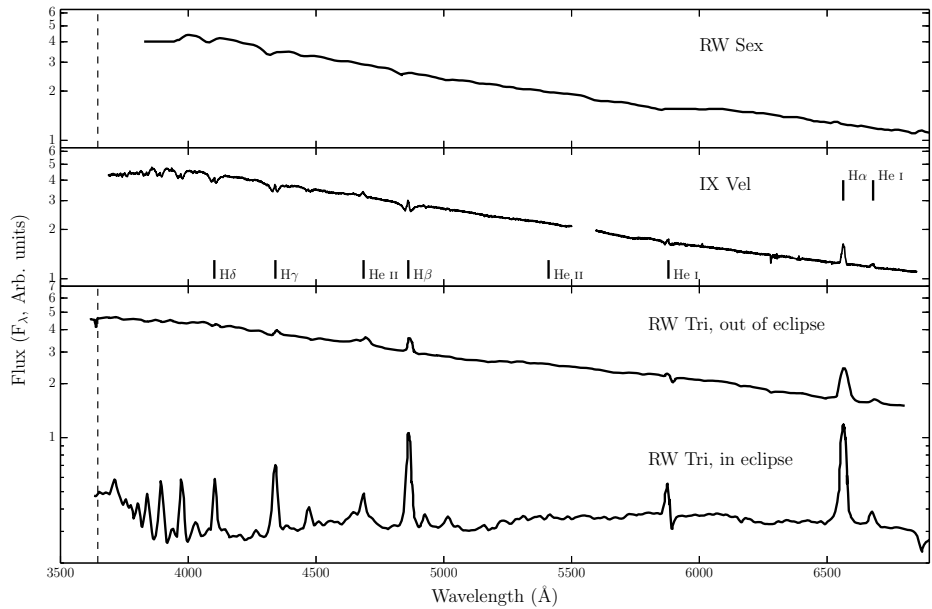}
\caption{
Optical spectra of three nova-like variables: 
RW Sex (top; Beuermann et al. 1992),
IX Vel (top middle; A. F. Pala \& B. T. Gaensicke, private communication) 
and RW Tri in and out of eclipse (bottom two panels; Groot et al. 2004).
The data for RW Sex and RW Tri were digitized from the respective publications,
and the IX Vel spectrum was obtained using the XSHOOTER spectrograph 
on the Very Large Telescope on 2014 October 10.
These systems have approximate inclinations of $30^\circ$, $60^\circ$ and $75^\circ$ 
(see section 5.4) respectively. 
The trend of increasing Balmer line emission with inclination can be seen.
In RW Tri strong single-peaked emission in the Balmer lines is seen even
in eclipse, indicating that the lines may be formed in a spatially
extensive disk wind, and there is even a suggestion 
of a (potentially wind-formed) recombination continuum in the eclipsed
spectrum. We have attempted to show each spectrum over a similar dynamic range.
}
\label{novalikes}
\end{figure*}

Cataclysmic variables (CVs) are systems in which a white dwarf
accretes matter from a donor star via Roche-lobe overflow. In
non-magnetic systems this accretion is mediated by a Keplerian disk
around the white dwarf (WD). Nova-like variables (NLs) are a subclass
of CVs in which the  disk is always in a relatively
high-accretion-rate state ($\dot{M} \sim
10^{-8}$~M$_{\odot}$~yr$^{-1}$).  This makes NLs an excellent
laboratory for studying the properties of steady-state accretion
disks.

It has been known for a long time that winds emanating from the
accretion disk are important in shaping the ultraviolet (UV) spectra
of high-state CVs \citep{heap1978, greensteinoke1982}. The most spectacular evidence for such
outflows are the P-Cygni-like profiles seen in UV resonance lines such as
\civfull\ (see e.g. Cordova \& Mason
1982\nocite{cordova1982}). Considerable effort has been spent over the
years on understanding and modelling these UV features (e.g. Drew \&
Verbunt 1985\nocite{drewverbunt1985}; Mauche \& Raymond
1987\nocite{maucheraymond1987}; Drew 1987; Shlosman \& Vitello 1993; [hereafter
SV93]\nocite{SV93}; Knigge, Woods \& Drew 1995\nocite{KWD95}; 
Knigge \& Drew 1997\nocite{kd1997}; 
Knigge et al. 1997\nocite{knigge1997}; Long \& Knigge 2002 [hereafter LK02]\nocite{LK02}, 
Noebauer et al. 2010\nocite{noebauer};
Puebla et al. 2011\nocite{puebla2011}). The basic picture emerging from these efforts is
of a slowly accelerating, moderately collimated bipolar
outflow that carries away $\simeq 1\% - 10\%$ of the accreting
material. State-of-the-art simulations of line formation in this type
of disk wind can produce UV line profiles that are remarkably similar
to observations.

Much less is known about the effect of these outflows on the optical
spectra of high-state CVs. These spectra are typically characterized
by H and He emission lines superposed on a blue continuum. In many
cases, and particularly in the SW~Sex subclass of NLs
\citep{HSK86,DR95}, these lines are single-peaked. This is contrary to
theoretical expectations for lines formed in accretion disks, which
are predicted to be double-peaked \citep{smak1981, hornemarsh1986}. 
{\em Low-state} CVs (dwarf novae in quiescence) do, in fact,
exhibit such double-peaked lines \citep{marshhorne1990}. 

Murray \& Chiang (1996, 1997; hereafter referred to collectively as MC96)\nocite{MC96, MC97} 
have shown that the presence of disk winds may
offer a natural explanation for the single-peaked optical emission lines in
high-state CVs, since they can strongly affect the radiative transfer
of line photons. Strong support for a significant wind contribution to the
optical emission lines comes from observations of eclipsing
systems. There, the single-peaked lines are often only weakly
eclipsed, and a significant fraction of the line flux remains visible
even near mid-eclipse \citep[e.g.][]{baptista2000,groot2004}. 
This points to line formation in a spatially
extended region, such as a disk wind (see Fig.~\ref{novalikes}).
Further evidence for a wind contribution to the optical lines comes
from isolated observations of P-Cygni-like line profiles even in optical
lines, such as \ha\ and He \textsc{i} $\lambda5876$ \citep{patterson1996, RN98, kafka2004}.

Could disk winds also have an impact on the UV/optical {\em continuum}
of high-state CVs? This continuum is usually thought to be dominated
by the accretion disk and modelled by splitting the disk into
a set of concentric, optically thick, non-interacting annuli following
the standard $T_{eff}(R) \propto R^{-3/4}$ radial temperature
distribution \citep{shakurasunyaev1973}. In such
models, each annulus is taken to emit either as a blackbody or,
perhaps more realistically, as a stellar/disk atmosphere model
\citep{Schwarzenberg-Czerny1977,wade1984,wade1988}.
In the latter case, the local surface gravity, $\log{g}(R)$, is
assumed to be set solely by the accreting WD, since self-gravity is
negligible in CV disks.

Attempts to fit the observed spectral energy distributions (SEDs) of
high-state CVs with such models have met with mixed success. In
particular, the SEDs predicted by most stellar/disk atmosphere models 
are too blue in the UV \citep{wade1988,long1991,long1994,knigge1998} and exhibit
stronger-than-observed Balmer jumps in absorption 
\citep{wade1984,haug1987,ladous1989b,knigge1998}. One possible
explanation for these problems is that these models fail to capture
all of the relevant physics. Indeed, it has been argued that a
self-consistent treatment can 
produce better agreement with observational data (e.g. Shaviv et
al. 1991;  but see also Idan et al. 2010).
\nocite{idanshaviv2010} \nocite{shaviv1991}
However, an alternative explanation, suggested by Knigge et al.
(1998b; see also Hassall et al. 1985)\nocite{KLWB98,hassall}, 
is that recombination continuum emission from the base of the 
disk wind might fill in the disk's
Balmer absorption edge and flatten the UV spectrum. 

\nocite{groot2004}
\nocite{beuermann1990}
\nocite{beuermann1992}
\nocite{higginbottom2013}

Here, we carry out Monte Carlo radiative transfer simulations in
order to assess the likely impact of accretion disk winds on the
optical spectra of high-state CVs. More specifically, our goal is to
test whether disk winds of the type developed to account for the UV
resonance lines would also naturally produce significant amounts of  
optical line and/or continuum emission. In order to achieve this, we
have implemented the `macro-atom' approach developed by Lucy
(2002, 2003) into the Monte Carlo ionization and radiative transfer
code described by LK02 (a process initiated by Sim et al. 2005; hereafter SDL05). 
With this upgrade, the code is able to deal correctly with processes involving
excited levels, such as the recombination emission produced by CV
winds. 

The remainder of this paper is organized as follows. In Section~2, we
briefly describe the code and and the newly implemented macro-atom
approach. In Section~3, we describe the kinematics and geometry of our
disk wind model. 
In Section~4, we present spectra simulated from the benchmark model
employed by LK02, and, in Section~5, we present a revised model
optimized for the optical waveband. In Section~6, we summarize our
findings.

%
%

\section{\sc{python}: A Monte Carlo Ionization and Radiative Transfer Code}

\py\ is a Monte Carlo  ionization and radiative transfer code which
uses the Sobolev approximation to treat line transfer 
\citep[e.g.][]{sobolev1957,sobolev1960,rybickihummer1978}. 
The code has already been described extensively by LK02, SDL05 and Higginbottom et al. (2013; hereafter H13), so here we provide only a brief summary of its operation, 
focusing particularly on new
aspects of our implementation of macro-atoms into the code. 

\subsection{Basics} 

\py\ operates in two distinct stages. First, the ionization state,
level populations and temperature structure are calculated. This is
done iteratively, by 
propagating several populations of Monte Carlo energy quanta (`photons')
through a model wind. The geometric and kinemetic properties of the
outflow are specified on a pre-defined spatial grid. In each of these
iterations (`ionization cycles'), the code records estimators that 
characterize the radiation field in each grid cell. At the end 
of each ionization cycle, a new electron temperature is calculated
that more closely balances heating and cooling in the 
plasma. The radiative estimators and updated electron
temperature are then used to revise the ionization state of the wind,
and a new ionization cycle is started. The process is repeated until
heating and cooling are balanced throughout the wind. 

This converged model is then used as the basis for the second set of
iterations (`spectral cycles'). In these, the emergent spectrum over
the desired spectral range is synthesized by tracking populations of
energy packets through the wind and computing the emergent spectra at
a number of user-specified viewing angles.  

\py\ is designed to operate in a number of different
regimes, both in terms of the scale of the system and in terms of the
characteristics of the underlying radiation field.
It was originally developed by LK02 in order to model the UV spectra
of CVs with a simple biconical disk wind model. SDL05
\nocite{simmacro2005} used the code to model Brackett
and Pfund line profiles of H in young-stellar objects (YSOs). As part
of this effort, they implemented a `macro-atom' mode (see below) in
order to correctly treat H recombination lines with
\py. Finally, H13 used \py\ to model broad absorption line (BAL) QSOs. For
this application, an improved treatment of ionization was implemented,
so that the code is now capable of dealing with arbitrary
photo-ionizing SEDs, including non-thermal and multi-component ones. 

\subsection{Ionization and Excitation: `Simple Atoms'}
\label{simpleatoms}

Prior to SDL05, the relative ionization fractions for all atomic
species were estimated via the modified Saha equation (Mazzali \&
Lucy 1993)  
\begin{equation}
\frac{n_{j+1} n_e}{n_j} = W [\xi + W(1-\xi)]
\left(\frac{T_e}{T_R}\right)^{1/2}
\left(\frac{n_{j+1}n_e}{n_j}\right)^*_{T_R}. \label{ionization}
\end{equation}
Here, the `starred' term on the right represents abundances computed with
the Saha equation at temperature $T_R$, but using partition functions
from the dilute blackbody approximation. 
$W$ is an effective dilution factor, $\xi$ is the
fraction of recombinations going directly to the ground state, and
$T_R$ and $T_e$ are the radiation and electron temperatures,
respectively. This simple ionization scheme produces reasonable
results when the photoionizing SED can be approximated by a dilute
blackbody. This is the case for high-state CVs. (As noted above, an
improved, but more complex treatment of ionization that is appropriate
for more complex SEDs is described in H13.) 

Similarly, the relative excitation fractions within each ionization
stage of a given species were estimated via a modified (dilute) Boltzmann
equation,
\begin{equation}
\frac{n_{jk}}{n_j} = \frac{W g_k}{z_j(T_R)} \exp(-E_k/kT_R),
\end{equation}
where $n_{jk}$ is the population of level $k$ in ionic stage $j$,
$E_k$ is the energy difference between level $k$ and the ground state,
$g_k$ is the statistical weight of level $k$
and $z_j(T_R)$ is the partition function of ionic stage $j$. 

Finally, \py\ originally modelled all bound-bound processes as transitions
within a simple two-level atom \cite[e.g.][]{mihalas}. 
This framework was used for the treatment of line transfer and also
for the line heating and cooling calculations (see LK02). 
The approximation works reasonably well for resonance  
lines, such as \civfull, in which the lower level is the ground state.  
However, it is a poor approximation for many other
transitions, particularly those where the upper level
is primarily populated from above. Thus an improved method for
estimating excited level populations and simulating line transfer is
needed in order to model recombination lines and continua.

\subsection{Ionization and Excitation: Macro-Atoms}

Lucy (2002, 2003\nocite{lucy2002, lucy2003}; hereafter L02, L03) 
has shown that it is possible to calculate the emissivity of a gas in
statistical equilibrium accurately by quantising matter into
`macro-atoms', and radiant and kinetic energy into indivisible energy
packets (r- and k- packets, respectively). His macro-atom scheme
allows for all possible transition paths from a given level and
provides a full non-local thermodynamic equilibrium (NLTE) solution
for the level populations based on Monte Carlo estimators. The macro-atom
technique has already been used to model Wolf-Rayet star
winds \citep{sim2004}, AGN disk winds \citep{simlong2008, tatum2012},
supernovae \citep{kromersim2009, kerzendorfsim} and YSOs (SDL05). A full 
description of the approach can be found in L02 and L03. 

Briefly, macro-atom NLTE level populations and ionization fractions
are calculated by solving the statistical equilbrium equations between
each pair of levels. In the framework of the Sobolev escape probability formalism (Rybicki \& Hummer 1978; L02; Sim 2004), 
the bound-bound excitation rate, ${\cal R}_{lu}$, in an ion is given by 
\begin{equation}
{\cal R}_{lu} = B_{lu} n_l J_{est} + C_{lu} n_l n_e,
\end{equation}
where $u$ and $l$ denote the upper and lower levels, $C$ represents the
collisional rate coefficients, and $B$ is the usual Einstein
coefficient. $J_{est}$ is the Monte Carlo estimator for the mean intensity 
impinging on the Sobolev region, weighted by an angle-dependent escape probability, 
given by \citep{sim2004}
\begin{equation}
J_{est} = \frac{c}{4 \pi \nu_0 V} \sum_{i} w_i \frac{1 - e^{-\tau_{s,i}}}{\tau_{s,i}} \frac{1}{(dv/ds)_i}.
\end{equation}
Here $w$ is the photon weight (in luminosity units), $\nu_0$
is the line frequency, $dv/ds$ is the velocity gradient and
$\tau_s$ is the Sobolev optical depth.
The sum is over all photons that come into resonance with the line,
and thus represents an integral over solid angle.
The corresponding de-excitation rate is then 
\begin{equation}
{\cal R}_{ul} = \beta_{lu} A_{ul} n_u + B_{ul} n_u J_{est} +
C_{ul} n_u n_e,
\label{eq:nlte_rul}
\end{equation}
where $A$ is the usual Einstein coefficient. 
The quantity $\beta_{lu}$ is the {\em angle-averaged} probability 
that a given line photon will escape the Sobolev region.

In our implementation of the macro-atom approach, we also explicitly
take into account the photoionization and collisional ionization rates
between a lower level, $l$, and the continuum (or, in the case of ions
with more than one bound electron, the ground state of the upper ion),
$\kappa$,
\begin{equation}
{\cal R}_{l \kappa}= n_l \int_{\nu_0}^{\infty} \frac{ 4 \pi J_{\nu}
  \sigma_{\nu}}{h \nu} d\nu + C_{l \kappa} n_l n_e.
\end{equation}
Here, $\sigma_{\nu}$ is the photoionization cross section, and $J_{\nu}$
is the mean intensity. The corresponding recombination rate is given
by 
\begin{equation}
{\cal R}_{\kappa l} = \alpha_{\kappa l} n_{\kappa} n_e + C_{\kappa l}
n_\kappa n_e, \\
\end{equation}
where $\alpha_{\kappa l}$ is the radiative recombination coefficient
to level $l$. This treatment means that radiative and collisional
rates to and from all levels are considered when calculating both the
ionization state and the level populations, although we neglect 
ionization directly to excited levels of the upper ion. The
\cite{vanregemorter} approximation is used for collisional
transitions. This means that collisions between radiatively
forbidden transitions are not taken into account when one 
splits levels ito $l$- and $s$-subshells, as well
as principal quantum number, $n$ (as we have done with He~\textsc{i}; 
see section~\ref{sec:data}). Although this approximation is, in general, 
a poor one, the effect is second order in the physical 
regime where recombination lines are formed in our models. 
This is because bound-free processes are dominant in determining 
level populations and emissivities. We have verified that this 
is indeed the case in the He~\textsc{i} emission regions in our models.

\subsection{Ionization and Excitation: A Hybrid Approach}

SDL05 implemented a macro-atom treatment of H in \py\ and used
this to predict the observable properties of a pure H wind
model for YSOs. Our goal here is to simultaneously model the optical
and ultraviolet spectra of high-state CVs. Since the optical spectra
are dominated by H and He recombination lines, both of these species
need to be treated as macro-atoms. The UV spectra, on the other hand,
are dominated by resonance lines associated with metals. This means we
need to include these species in our models, but they can be treated 
with our (much faster) simple-atom approach. We have therefore
implemented a hybrid ionization and excitation scheme into \py. Any
atomic species can now be treated either in our simple-atom
approximation or with the full macro-atom machinery. In our CV
models, we treat H and He as macro-atoms and all metals as
simple-atoms. Species treated with either method
are full taken into account as sources of both bound-free opacity and line opacity, 
and contribute to the heating and cooling balance of the plasma.

\subsection{Atomic Data}
\label{sec:data}

We generally use the same atomic data as H13, which is an updated
version of that described by LK02. In addition, we follow SDL05 in
treating H as a 20-level atom, where each level is defined by
the principal quantum number, $n$. For the macro-atom treatment of
He, we have added the additional level and line information required 
from \textsc{Topbase} \citep{topbase2005}.  He~\textsc{ii} is treated
in much the same way as H, but with 10 levels. He~\textsc{i} has
larger energy differentials between different l-subshells and triplet
and singlet states. Thus, we still include levels up to $n=10$, but
explicitly treat the $l$ and $s$ sub-orbitals as distinct levels
instead of assuming they are perfectly `l-mixed'. This allows us
to model the singlet and  triplet He~\textsc{i} lines that are ubiquitous
in the optical spectra of CVs \citep[e.g.][]{dhillon1996}.

\subsection{Code Validation and Testing}

\begin{figure}
\centering
\includegraphics[width=0.5\textwidth]{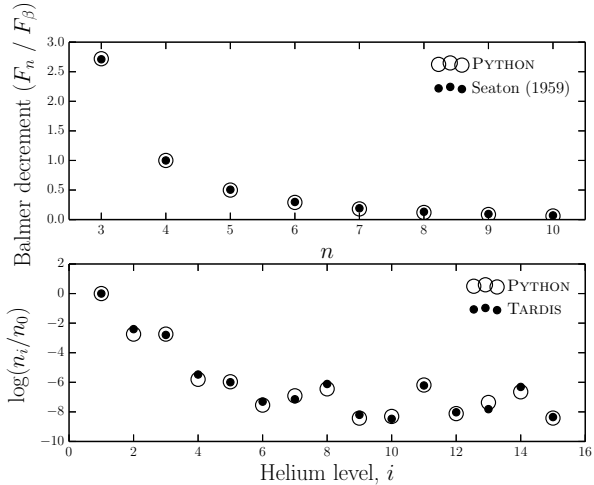}
\caption{
{\sl Top Panel:} `Case B' Balmer decrements computed 
with \textsc{Python} compared to analytic calculations
by Seaton (1959). Both calculations are calculated at $T_e=10,000$K.
(see Osterbrock 1989 for a discussion of this commonly used approximation).
{\sl Bottom Panel:}  a comparison of He I level populations (the most complex ion we currently 
treat as a macro-atom) between \py\ and \tar models. 
The calculation is conducted with physical parameters of $n_e=5.96\times10^4$~cm$^{-3}$,
$T_e=30,600$K, $T_R=43,482$K and $W=9.65\times10^{-5}$. 
Considering the two codes use different atomic data and \textsc{Tardis,} unlike \textsc{Python,} currently has a complete treatment of collisions between 
radiatively forbidden transitions, the factor of 
$<2$ agreement is encouraging. 
}
\label{tests}
\end{figure}

\py\ has been tested against a number of radiative transfer and
photoionization codes. LK02 and H13 conducted comparisons of 
ionization balance  with \cld \citep{cloudy2013}, demonstrating
excellent agreement. We have also carried out comparisons
of ionization and spectral synthesis with the supernova code
\textsc{Tardis.} \tar is described by 
\cite{kerzendorfsim}, and the spectral comparisons can be found
therein. For the effort reported here, we have additionally carried
out tests of the macro-atom scheme in \py. Fig.~\ref{tests} shows
two of these tests. In the top panel, we compare the Balmer series 
emissivities as predicted by \py\ in the l-mixed Case~B limit against the
analytical calculations by \cite{seaton1959}. In the bottom panel, we
compare \py\ and \tar predictions of  He \textsc{i} level populations
for a particular test case. Agreement is excellent for both H and He.

\nocite{osterbrock}
\nocite{seaton1959}

%
%

\section{Describing the System and its Outflow}

\py\ includes several different kinematic models of accretion disk
winds, as well as different options for describing the physical and
radiative properties of the wind-driving system under
consideration. Most of these features have already been discussed by
LK02 and H13, so below we only briefly recount the key aspects of the
particular system and wind model used in the present study.

\subsection{Wind Geometry and Kinematics}
\label{kinematics}

\begin{figure} 
\centering
\includegraphics[width=0.45\textwidth]{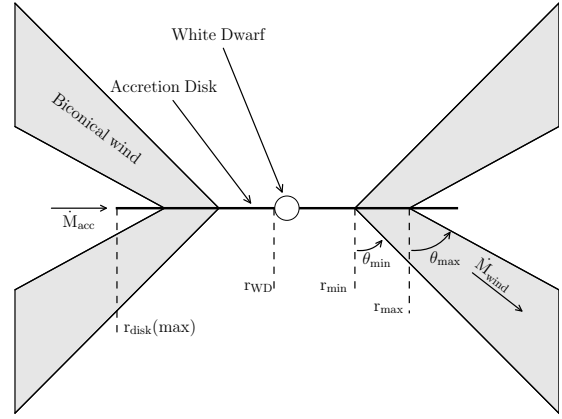}
\caption{Cartoon illustrating the geometry and kinematics of the benchmark CV wind model.}
\label{cartoon}
\end{figure}

We adopt the kinematic disk wind model developed by SV93. 
A schematic of this model is shown in
Fig.~\ref{cartoon}. In this parametrization, a smooth, biconical
disk wind emanates from the accretion disk between radii $r_{min}$ and 
$r_{max}$. The covering fraction of the outflow is also controlled by the
inner and outer opening angles of the wind, $\theta_{min}$ and
$\theta_{max}$, and the launch angle of the other streamlines is given
by 
\begin{equation}
\theta(r_0) = \theta_{min} + (\theta_{max} - \theta_{min}) \left(\frac{r_0 - r_{min}}{r_{max} - r_{min}} \right)^{\gamma},
\label{theta}
\end{equation}
where $r_0$ is the launch radius of the streamline.

\begin{figure}
\centering
\includegraphics[width=0.45\textwidth]{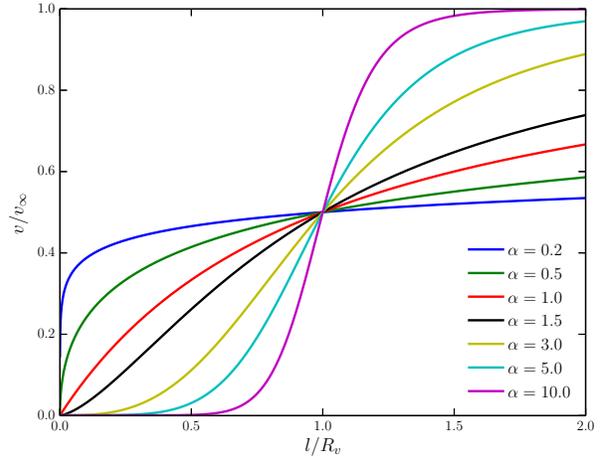}
\caption{
The adopted poloidal velocity law for various values of the
acceleration exponent, $\alpha$.
} 
\label{acc_law}
\end{figure}

The poloidal (non-rotational) velocity field of the wind, $v_l$, is given by
\begin{equation}
v_l=v_0+\left[v_{\infty}(r_0)-v_0\right]\frac{\left(l/R_v\right)^{\alpha}}{\left(l/R_v\right)^{\alpha}+1},
\label{v_law}
\end{equation}
where $l$ is the poloidal distance along a particular wind
streamline. The terminal velocity along a streamline, $v_{\infty}$, is
set to a fixed multiple of $v_{esc}$, the escape velocity at the launch
point. The launch velocity from the disk surface, $v_0$, is assumed to
be constant (set to $6$~km~s$^{-1}$). Once the wind is launched, it
accelerates, reaching half of its terminal velocity at $l = R_v$. The
velocity law exponent $\alpha$ controls how quickly the wind
accelerates. Larger values of $\alpha$ cause the main region of 
acceleration to occur close to $R_v$, whereas smaller values
correspond to fast acceleration close to the disk (see
Fig.~\ref{acc_law}). The rotational velocity $v_\phi$ is 
Keplerian at the base of the streamline 
and we assume conservation of specific angular momentum,
such that
\begin{equation}
v_\phi r = v_{k} r_0,
\label{v_law}
\end{equation}
where $v_{k}=(GM_{WD}/r_0)^{1/2}$.

The density at position $(r,z)$ in the wind, $\rho(r,z)$, is
calculated from the mass continuity equation, yielding
\begin{equation}
\rho(r,z) = \frac{r_0}{r} \frac{dr_0}{dr} \frac{\phi(r_0)}{v_z(r,z)}.
\label{density}
\end{equation}
Here, 
$v_z$ is the vertical velocity component and, following SV93,
$\phi(r_0)$ is the local mass-loss rate per unit area at $r_0$, 
defined as
\begin{equation}
\phi(r_0) \propto \dot{M}_{wind} r_0^\lambda \cos [\theta(r_0)].
\label{density}
\end{equation}
We adopt $\lambda = 0$ and normalize $\phi(r_0)$ by 
matching its integral over both sides of the disk
to the user-specified total mass-loss rate, $\dot{M}_{wind}$.

\subsection{Sources and Sinks of Radiation}
\label{radsources}

The net photon sources in our CV model are the accretion disk, the
WD and, in principle, a boundary layer with user-defined temperature
and luminosity. All of these radiating bodies are taken to be
optically thick, and photons striking them are assumed to be destroyed
instantaneously. The secondary star is not included as a radiation
source, but is included as an occulting body. This allows us to model
eclipses. Finally, emission from the wind itself is also accounted for, but
note that we assume the outflow to be in radiative equilibrium. Thus all
of the heating of the wind, as well as its emission, is ultimately
powered by the radiation field of the net photon sources in the
simulation. In the following sections, we will describe our treatment
of these system components in slightly more detail.

\subsubsection{Accretion Disk}

\py\ has some flexibility when treating the accretion 
disk as a source of photons. The disk is broken down into annuli 
such that each annulus contributes an equal amount to the bolometric
luminosity. We take the disk to be geometrically thin, but optically
thick, and thus adopt the temperature profile of a standard
\cite{shakurasunyaev1973} $\alpha$-disk. An annulus can then
be treated either as a blackbody with the corresponding effective
temperature or as a stellar atmosphere model with the appropriate
surface gravity and effective temperature. Here, we use blackbodies 
during the ionization cycles and to compute our Monte Carlo
estimators. However, during the spectral synthesis stage of the 
simulation we use stellar atmosphere models. This produces more
realistic model spectra and allows us to test if recombination
emission from the wind base can fill in the Balmer jump, which is
always in absorption in these models. Our synthetic stellar atmosphere
spectra are calculated with
\textsc{Synspec}\footnote{http://nova.astro.umd.edu/Synspec43/synspec.html}
from either Kurucz \citep{kurucz1991} atmospheres (for $T_{eff} \leq
50,000$~K) or from \textsc{TLUSTY} \citep{tlusty} models (for $T_{eff} > 50,000$~K). 

\subsubsection{White Dwarf}

The WD at the center of the disk is always present as a spherical occulting
body with radius $R_{WD}$ in \py\ CV models, but it can also be included
as a source of radiation. In the models presented here, we treat the
WD as a blackbody radiator with temperature $T_{WD}$ and luminosity
$L_{WD} = 4\pi R_{WD}^2 \sigma T_{WD}^4$. 

\subsubsection{Boundary Layer}

It is possible to include radiation from a boundary layer (BL) between
the disk and the WD. In \py, the BL is described as
a blackbody with a user-specified effective temperature and
luminosity. In the models presented here, we have followed LK02 in setting
the BL luminosity to zero. However, we have confirmed that the addition of an isotropic
BL with $L_{BL} = 0.5 L_{acc}$ and temperatures in the range $80~{\rm
kK} \leq T_{BL} \leq 200~{\rm kK}$ would not change any of our main
conclusions. 

\subsubsection{Secondary Star}

The donor star is included in the system as a pure radiation sink, 
i.e. it does not emit photons, but absorbs any photons that strike its
surface. The secondary is assumed to be Roche-lobe filling, so its
shape and relative size are defined by setting the mass ratio of the system, 
$q = M_2/M_{WD}$. The inclusion of the donor star as an occulting body
allows us to model eclipses of the disk and the wind. For this
purpose, we assume a circular orbit with a semi-major axis $a$ and 
specify orbital phase such that $\Phi_{orb} = 0$ is the
inferior conjunction of the secondary (i.e. mid-eclipse for $i \simeq
90^o$).

%
%

\begin{table}
\centering
\begin{tabular}{p{2cm}p{2cm}p{2cm}}
\multicolumn{2}{|l|}{Model Parameters}  \\
\hline Parameter 	&	 Model A  & Model B \\ 
\hline \hline 
$M_{WD}$ 	 &	 $0.8~M_{\odot}$  &     \\ 
$R_{WD}$ 	 &	 $7\times10^{8}$~cm  & \\ 
$T_{WD}$ 	 &	 $40,000$~K        &  \\
$M_{2}$ 	& -&	 $0.6~M_{\odot}$   \\ 
$q$ 	&- &	 $0.75$   \\ 
$P_{orb}$ 	&- &	 $5.57$~hr   \\ 
$a$ 	& -&	 $194.4~R_{WD}$   \\ 
$R_2$   &   -  &	 $69.0~R_{WD}$  \\ 
$\dot{M}_{acc}$ 	 &	 $10^{-8}~M_{\odot}yr^{-1}$  &\\ 
$\dot{M}_{wind}$  &	$10^{-9}~M_{\odot}yr^{-1}$å  & \\ 
$r_{min}$ 	&	 $4~R_{WD}$ &  \\ 
$r_{max}$ 	&	 $12~R_{WD}$  &  \\ 
$r_{disk}$(max) 	&	 $34.3~R_{WD}$  &  \\ 
$\theta_{min}$ 	&	 $20.0^{\circ}$  &  \\ 
$\theta_{max}$ 	&	 $65.0^{\circ}$  &  \\ 
$\gamma$ 	&	 $1$  &  \\ 
$v_{\infty}$ 	&	 $3~v_{esc}$  &  \\ 
$R_v$ 	        &	 $100~R_{WD}$  &  $142.9~R_{WD}$  \\ 
$\alpha$ 	&	 $1.5$   &   $4$\\
\end{tabular}
\centering
\caption{
Parameters used for the geometry and kinematics of the benchmark 
CV model (model A), which is optimized for the UV band, and a model
which is optimized for the optical band and described in section 5 (model B).
For model B, only parameters which are altered are given - otherwise the
model A parameter is used. $P_{orb}$ is the orbital period 
(the value for RW Tri from Walker 1963 is adopted, see section 5.4) and 
$R_2$ is the radius of a sphere with the volume of the secondary's Roche lobe. 
Other quantities are defined in the text or Fig.~\ref{cartoon}.
Secondary star parameters are only quoted for 
model B as we do not show eclipses with the 
benchmark model (see section 5.4).
}
\label{wind_param}
\label{modelb_table}
\end{table}

\nocite{walker1963}

\section{A Benchmark Disk Wind Model}
\label{modela}

Our main goal is to test whether the type of disk wind model that has
been successful in explaining the UV spectra of CVs could also have a
significant impact on the optical continuum and emission line spectra
of these systems. In order to set a benchmark, we therefore begin by
investigating one of the fiducial CV wind models that was used by SV93
and LK02 to simulate the UV spectrum of a typical high-state
system. The specific parameters for this model (model A) are listed in
Table~1. A key point is that the wind mass-loss rate in this model is
set to 10$\%$ of the accretion rate through the disk. The inner edge of
the wind ($r_{min}$) is set to $4~R_{WD}$ following SV93. 
The sensitivity to some of these parameters is briefly discussed in
section~5. 

\subsection{Physical Structure and Ionization State}
\label{modela_ionization}

\begin{figure*}
\includegraphics[width=0.8\textwidth]{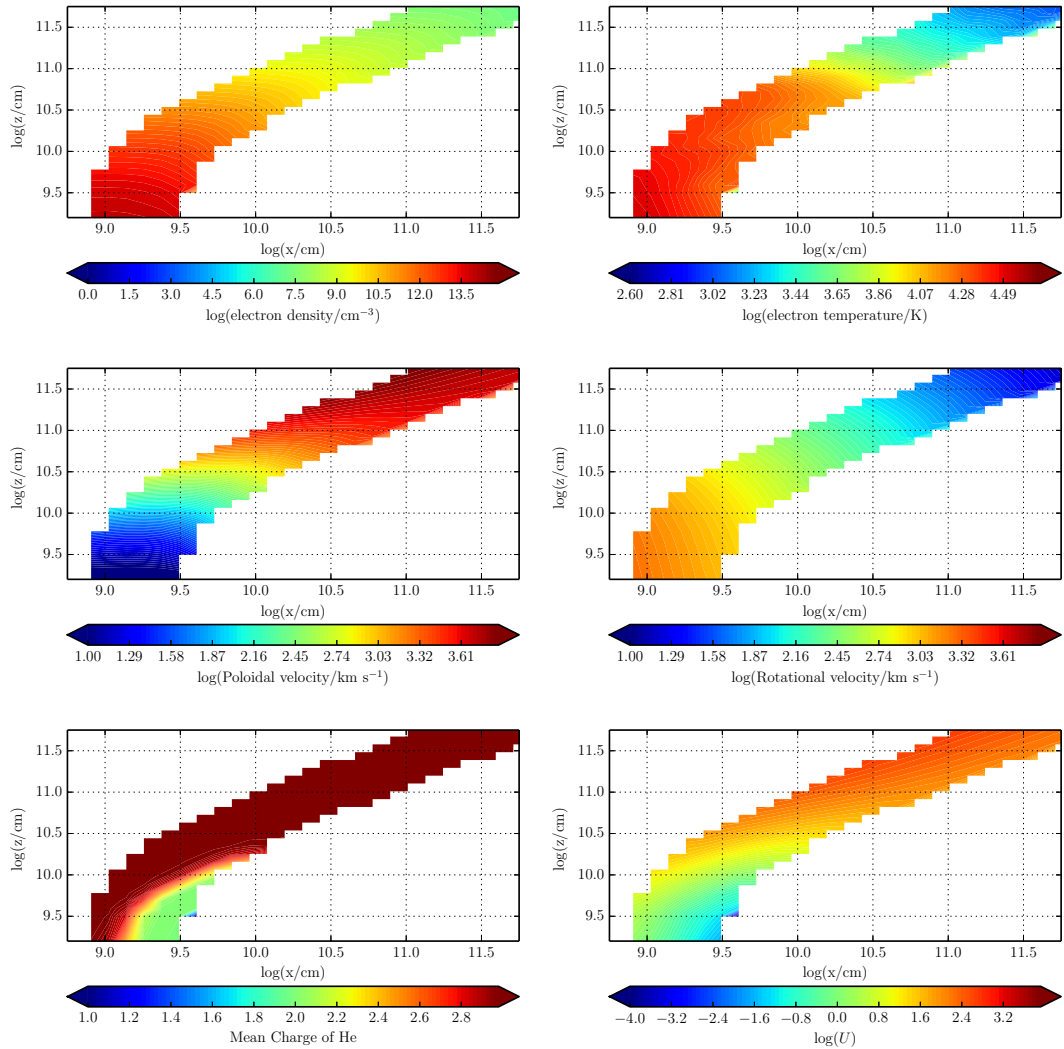}
\caption{
The physical properties of the wind -- note the logarithmic scale. 
Near the disk plane the wind is dense, with low poloidal velocities.
As the wind accelerates it becomes less dense
and more highly ionized. The dominant He ion
is almost always He III, apart from in a small
portion of the wind at the base, which is partially shielded
from the inner disk.
}
\label{wind}
\end{figure*}

Fig.~\ref{wind} shows the physical and ionization structure 
of the benchmark disk wind model. The ionization parameter shown in the bottom
right panel is given by

\begin{equation}
U = \frac{4\pi}{n_H c}\int_{13.6{\rm{eV}}}^{\infty}\frac{{J_{\nu}d\nu}}{h\nu},
\end{equation}
 
\noindent where $\rm{n_H}$ is the local number density of H, and $\nu$ denotes photon 
frequency. The ionization parameter is a useful measure of the ionization state of a plasma, 
as it evaluates the ratio of the number density of ionizing photons to the local 
H density.

There is an obvious drop-off in density
and temperature with distance away from the disk, so any line
formation process that scales as $\rho^2$ -- i.e. recombination and
collisionally excited emission -- should be expected to operate
primarily in the dense base of the outflow. Moreover, a comparison of
the rotational and poloidal velocity fields shows that rotation
dominates in the near-disk regime, while outflow dominates further out
in the wind. 

The ionization equation used in the `simple atom' approach used by
LK02 (see section~\ref{simpleatoms}) should be a reasonable approximation to
the photoionization equilibrium in the benchmark wind model. Even
though the macro-atom treatment of H and He does affect the 
computation of the overall ionization equilibrium, we would expect the
resulting ionization state of the wind to be quite similar to that
found by LK02. The bottom panels in Fig.~\ref{wind} confirm that this
is the case. In particular, He is fully ionized
throughout most of the outflow, except for a small region near the
base of the wind, which is shielded from the photons produced by the
hot inner disk. In line with the results of LK02, we also find
that C\textsc{iv} is the dominant C ion throughout the wind,
resulting in a substantial absorbing column across a large range of
velocities. As we shall see, this produces the broad, deep and
blue-shifted C\textsc{iv}~$\lambda1550$ absorption line that
is usually the most prominent wind-formed feature in the UV spectra of
low-inclination nova-like CVs.

\begin{figure*}
\includegraphics[width=0.9\textwidth]{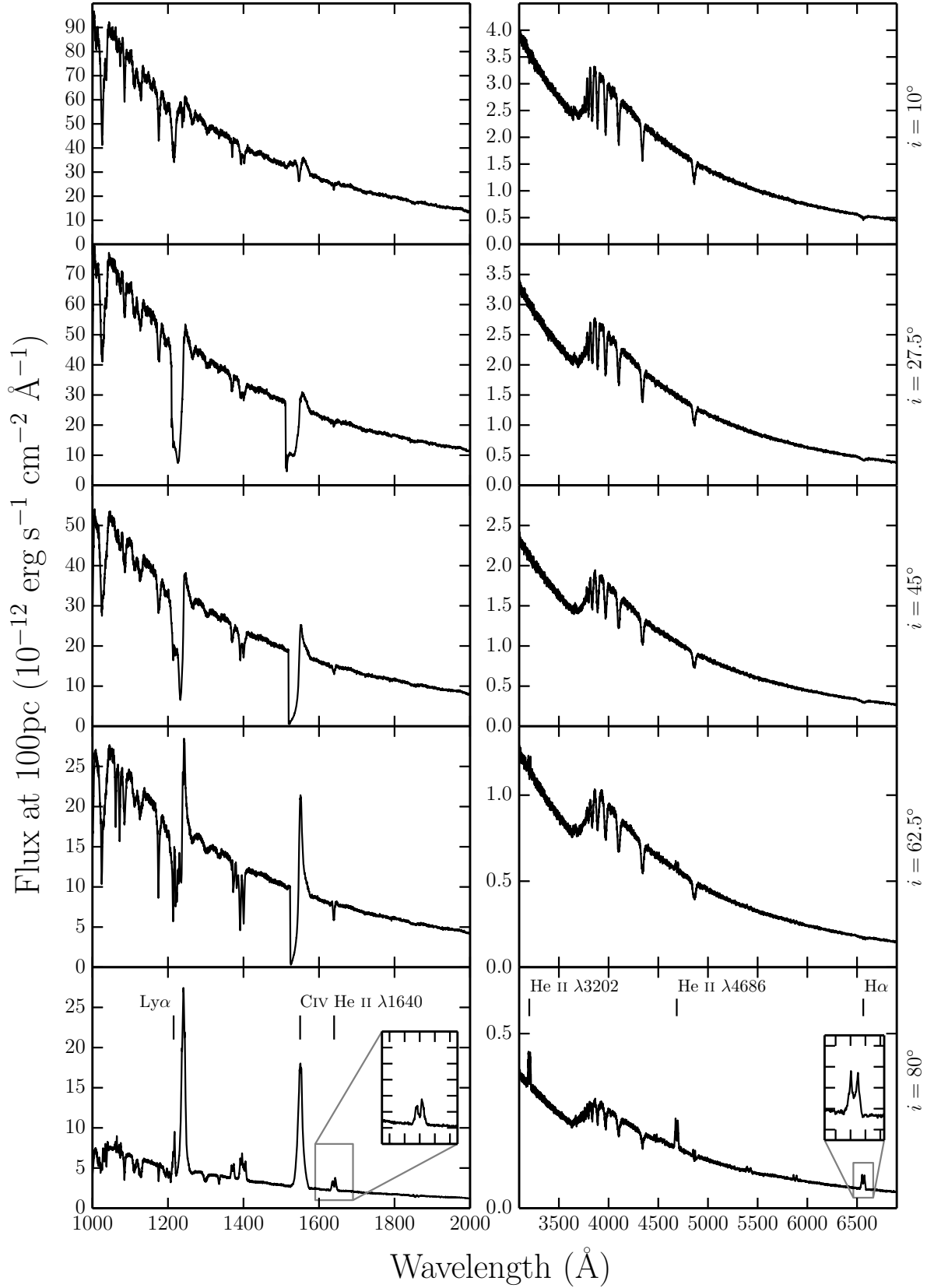}
\caption{
UV (left) and optical (right) synthetic spectra for model A, our benchmark model,
computed at sightlines of 10, 27.5, 45, 62.5 and 80 degrees.	
The inset plots show zoomed-in line profiles for 
\heiiuv\ and \ha. Double-peaked line emission can be seen in 
\heiiuv, \heiiopt, \ha\ and some He I lines, but the 
line emission is not always sufficient to overcome the absorption
cores from the stellar atmosphere models. The model
also produces a prominent \heiioptnew\ line at high inclinations.
}
\label{spec}
\end{figure*}

\subsection{Synthetic Spectra}
\label{modela_spectrum}

We begin by verifying that the benchmark model still produces UV
spectra that resemble those observed in CVs. We do expect this to be
the case, since the ionization state of the wind has not changed
significantly from that computed by LK02 (see section~\ref{modela_ionization}). 
The left column of panels in Fig.~\ref{spec} shows that this expectation
is met: all of the strong metal resonance
lines -- notably N~\textsc{v}~$\lambda1240$,
Si~\textsc{iv}~$\lambda1400$ and C~\textsc{iv}~$\lambda1550$ -- 
are present and exhibit clear P-Cygni profiles
at intermediate inclinations. In addition, however, we now also find
that the wind produces significant Ly$\alpha$ and
He~\textsc{ii}~$\lambda1640$ emission lines. 

\begin{figure} 
\includegraphics[width=0.45\textwidth]{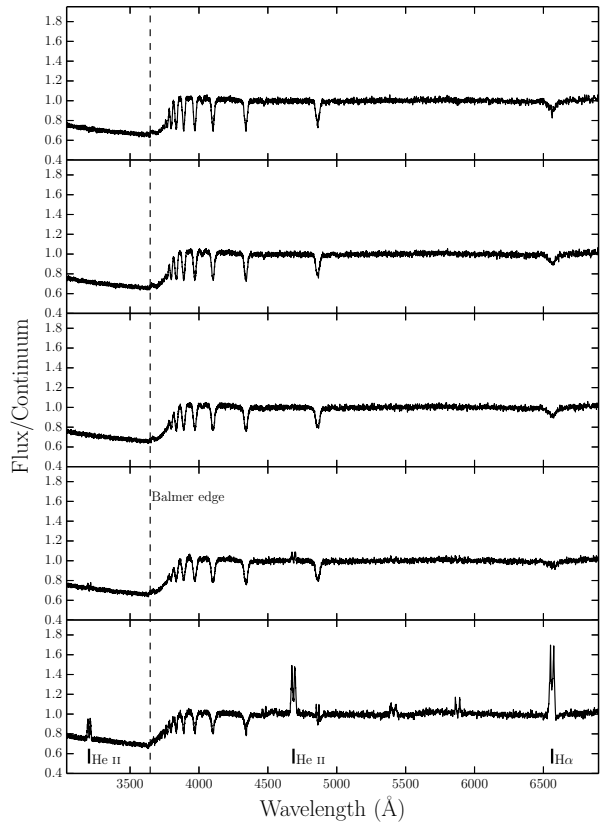}
\caption{Synthetic optical spectra from model A computed for 
sightlines of 10, 27.5, 45, 62.5 and 80 degrees. In these plots
the flux is divided by a polynomial fit to the 
underlying continuum redward of the Balmer edge, so that 
line-to-continuum ratios and the true depth of the
Balmer jump can be shown.}
\label{spec_continuum}
\end{figure}

Fig.~\ref{spec} (right-hand panel) and Fig.~\ref{spec_continuum}
show the corresponding optical spectra produced for
the benchmark model, and these do exhibit some emission lines
associated with H and He. We see a general trend from absorption lines to emission lines 
with increasing inclination, as one might expect from our wind
geometry. This trend is consistent with observations, as can be seen
in Fig.~1. However, it is clear that this particular model
does not produce all of the lines seen in observations of high-state CVs.
The higher-order Balmer series lines are too weak
to overcome the intrinsic absorption from the disk atmosphere, and the wind 
fails to produce any observable emission at low and intermediate inclinations.
This contrasts with the fact that emission lines are seen 
in the optical spectra of (for example) V3885 Sgr \citep{hartley2005}
and IX Vel \citep[][see also Fig.~1]{beuermann1990}.

The emissivity of these recombination 
features scales as $\rho^2$, meaning that they form almost entirely in the 
dense base of the wind, just above the accretion disk. Here, the
velocity field of the wind is still dominated by rotation, rather than
outflow, which accounts for the double-peaked shape of the lines. In
principle, lines formed in this region can still be single peaked,
since the existence of a poloidal velocity {\em gradient} changes the
local escape probabilities (MC96). However, as
discussed further in section~5.3, the 
radial velocity shear in our
models is not high enough for this radiative transfer effect
to dominate the line shapes.

\begin{figure} 
\includegraphics[width=0.45\textwidth]{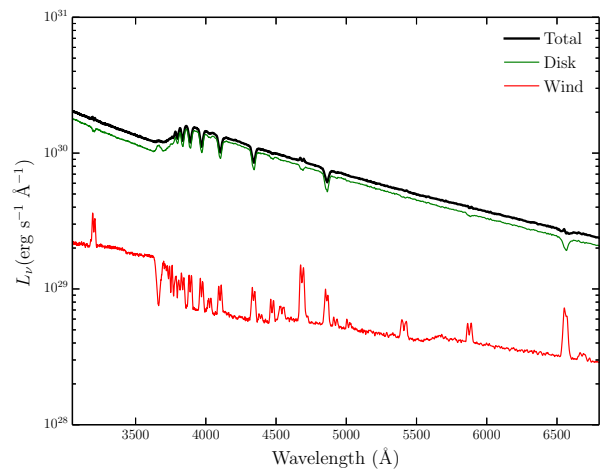}
\caption{Total packet-binned spectra across all viewing angles, in units
of monochromatic luminosity.
The thick black line shows the total 
integrated escaping spectrum, 
while the green line shows disk photons which escape without being reprocessed by
the wind. The red line show the contributions from reprocessed photons. 
Recombination continuum emission blueward of the Balmer 
edge is already prominent relative to other wind continuum processes, but is not sufficient
to fill in the Balmer jump in this specific model}
\label{cont}
\end{figure}

The Balmer jump is in absorption at all inclinations for our benchmark
model. This is due to the stellar atmospheres we have used to
model the disk spectrum; it is not a result of photoabsorption in the
wind. In fact, the wind spectrum exhibits the Balmer jump in {\em
emission}, but this is not strong enough to overcome the intrinsic
absorption edge in the disk spectrum. This is illustrated in
Fig.~\ref{cont}, which shows the angle-integrated spectrum of the system,
i.e. the spectrum formed by all escaping photons, separated into the
disk and wind contributions. Even though the wind-formed Balmer
recombination continuum does not completely fill in the Balmer
absorption edge in this model, it does already contribute
significantly to the total spectrum. This suggests that modest changes 
to the outflow kinematics might boost the wind continuum and produce
emergent spectra with weak or absent Balmer absorption edges. 

\newpage

%
%

\section{A Revised Model Optimized for Optical Wavelengths}

The benchmark model discussed in section~\ref{modela} was originally
designed to reproduce the wind-formed lines seen in the UV spectra of
high-state CVs. As we have seen, this model does produce some observable
optical emission. We can now attempt to construct a model that more closely 
matches the observed optical spectra of CVs. 

Specifically, we aim to assess whether a revised model can:

\begin{itemize}
         \item account for all of the lines we see in optical spectra of CVs while preserving
the UV behaviour;
         \item produce single-peaked Balmer emission lines; 
         \item generate enough of a wind-formed recombination continuum
to completely fill in the disk's Balmer absorption edge for 
reasonable outflow parameters.
\end{itemize} 

The emission measure of a plasma is directly proportional to its density.
The simplest way to simultaneously affect the density in the wind (for fixed mass loss rate),
as well as the velocity gradients, is by modifying the poloidal velocity
law. Therefore, we focus on just two kinematic variables (section~\ref{kinematics}):

\begin{itemize}
         \item the acceleration length, $R_v$, which controls the
        distance over which the wind accelerates to $\frac{1}{2}~v_{\infty}$;
         \item the acceleration exponent, $\alpha$, which controls the rate 
         at which the poloidal velocity changes near $R_v$.
\end{itemize} 

The general behaviour we might expect is that outflows with denser
regions near the wind base -- i.e. winds with larger $R_{v}$ and/or
larger $\alpha$ -- will produce stronger optical emission signatures. 
However, this behaviour may be moderated by the effect of the increasing
optical depth through this region, which can also affect the line profile shapes. 
In addition, modifying $R_v$ also increases the emission {\em volume}.
Based on a preliminary exploration of models with different kinematics,
we adopt the parameters listed in table~\ref{modelb_table}
for our `optically optimized' model (model B). 


\subsection{Synthetic Spectra}

\begin{figure*}
\includegraphics[width=0.9\textwidth]{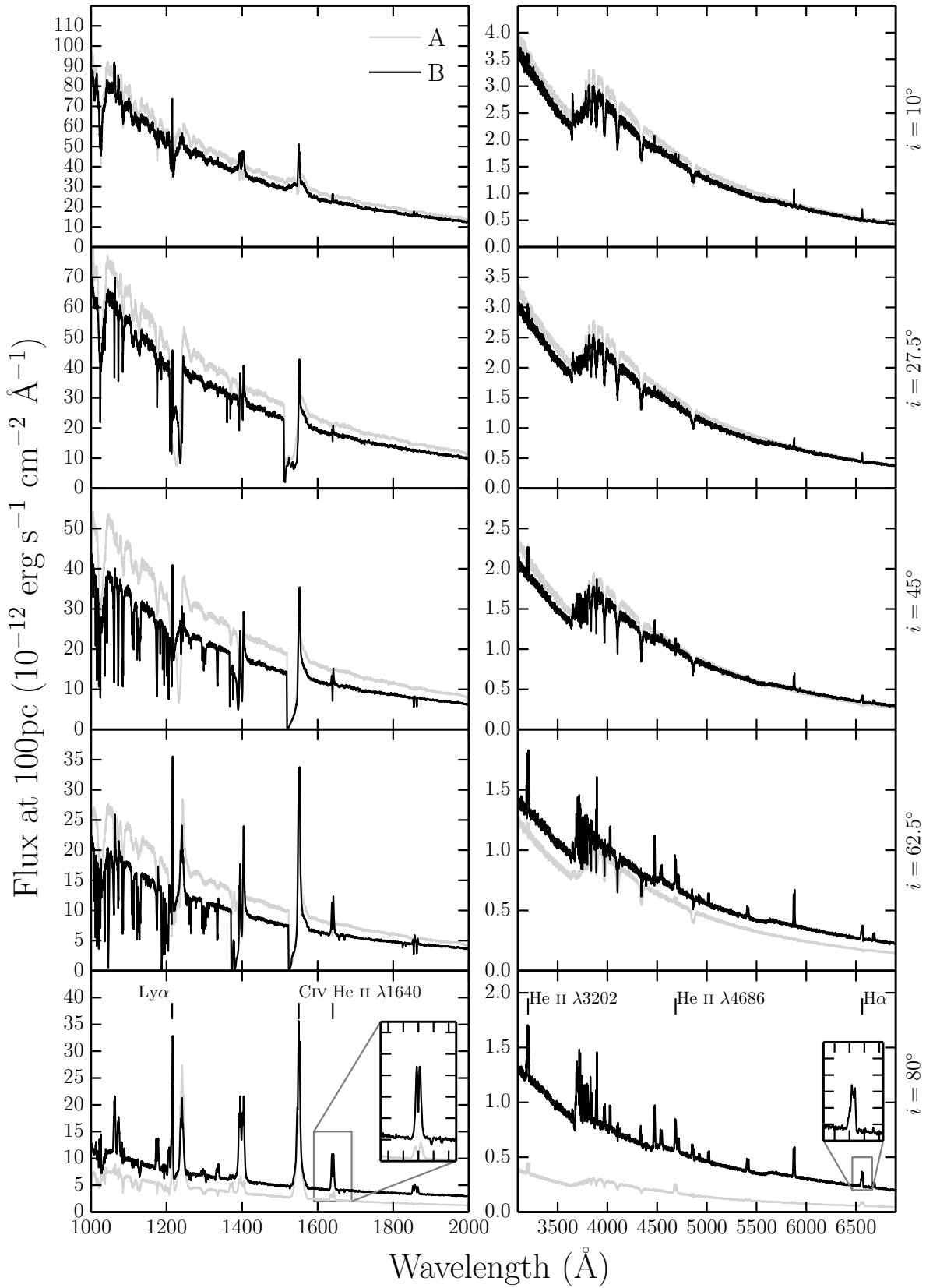}
\caption{
UV (left) and optical (right) synthetic spectra for model B computed at
sightlines of 10, 27.5, 45, 62.5 and 80 degrees. 
Model A is shown in grey for comparison.	
The inset plots show zoomed-in line profiles for 
\heiiuv\ and \ha. The Balmer and He
are double-peaked, albeit with narrower profiles.
Strong \heiiopt\ emission can be seen, as well as a trend
of a deeper Balmer jump with decreasing inclination.
}
\label{uvoptb}
\end{figure*}

Fig.~\ref{uvoptb} shows the UV and optical spectra for the
optically optimized model for the full range of inclinations. 
As expected, the trend from absorption to emission 
in the optical is again present, but in this revised model we produce emission
lines in the entire Balmer series at high inclinations, as well as the observed lines 
in He~\textsc{ii} and He~\textsc{i}. This can be seen more clearly in the 
continuum-normalized spectrum in Fig.~\ref{continuumb}.

Two other features are worth noting in the optical
spectrum. First, the collisionally excited Ca~{\sc ii} emission line at 3934~\AA\ 
becomes quite prominent in our densest models. Second, our model predicts a detectable
He~\textsc{ii} recombination line at 3202~\AA. This is the He
equivalent of Paschen~$\beta$ and should be expected in all systems that
feature a strong He~\textsc{ii}~$\lambda4686$ line (the He
equivalent of Paschen~$\alpha$). 
This line is somewhat unfamiliar observationally, because it 
lies bluewards of the atmospheric cut-off, but
also redwards of most ultraviolet spectra. 

Our models do not exhibit P-Cygni profiles in the optical lines.
This is perhaps not surprising. LK02 and SV93 originally designed such models
to reproduce the UV line profiles. Thus, most of the wind
has an ionization parameter of $\log U \sim 2$ (see Fig.~\ref{wind}).
This means H and He are fully ionized throughout 
much of the wind and are successful in producing recombination features.
However, the line opacity throughout the wind is too
low to produce noticeable blue shifted absorption. 
We suspect that the systems that exhibit such profiles must 
possess a higher degree of ionization stratification, although the lack 
of contemporary observations means it is not known for certain if the 
P-Cygni profiles in UV resonance lines and optical H and He lines exist simultaneously.
Ionization stratification could be caused by a clumpy flow, in which the ionization state 
changes due to small scale density fluctuations, or a stratification in density
and ionizing radiation field over larger scales.
Invoking clumpiness in these outflows is not an unreasonable
hypothesis. Theories of line-driven winds predict an unstable flow
\citep{macgregor1979,owockirybicki1984,owockirybicki1985}, and
simulations of CV disk winds also produce density inhomogeneities 
\citep{proga1998,pkdh2002}.
Tentative evidence for clumping being directly related to P-Cygni optical lines
comes from the fact that \cite{prinja2000}
found the dwarf nova BZ Cam's outflow to be unsteady and highly mass-loaded in outburst,
based on observations of the UV resonance lines.
This system has also exhibited P-Cygni profiles in He~\textsc{i}~$\lambda5876$
and \ha\ when in a high-state \citep{patterson1996,RN98}. 
The degree of ionization and density variation and 
subsequent line opacities may be affected by our model parameters
and the specific parameterisation we have adopted.

In the UV, the model still produces all the observed lines, 
and deep P-Cygni profiles are produced in the normal resonance lines,
as discussed in section 4.2. However, the UV spectra also
display what is perhaps the biggest problem with this revised model,
namely the strength of resonance line emission 
at low and intermediate inclinations.
In order to generate strong optical wind signatures, we have adopted wind
parameters that lead to very high densities at the base of the wind
($n_e\sim10^{13}-10^{14}$~cm$^{-3}$). This produces
the desired optical recombination emission, but also increases the
role of collisional excitation in the formation of the UV resonance
lines. This explains the pronounced increase in the emission component 
of the C\textsc{iv} $\lambda1550$ resonance line, for example, relative to
what was seen in the benchmark model (compare Figures~\ref{spec} and
\ref{uvoptb}). The strength of this component in the revised model 
is probably somewhat too high to be consistent with UV observations 
of high-state CVs (see e.g. Long et al. 1991, 1994; Noebauer et al. 2010).
\nocite{long1991,long1994, noebauer}

\subsection{Continuum Shape and the Balmer Jump}

\begin{figure} 
\includegraphics[width=0.45\textwidth]{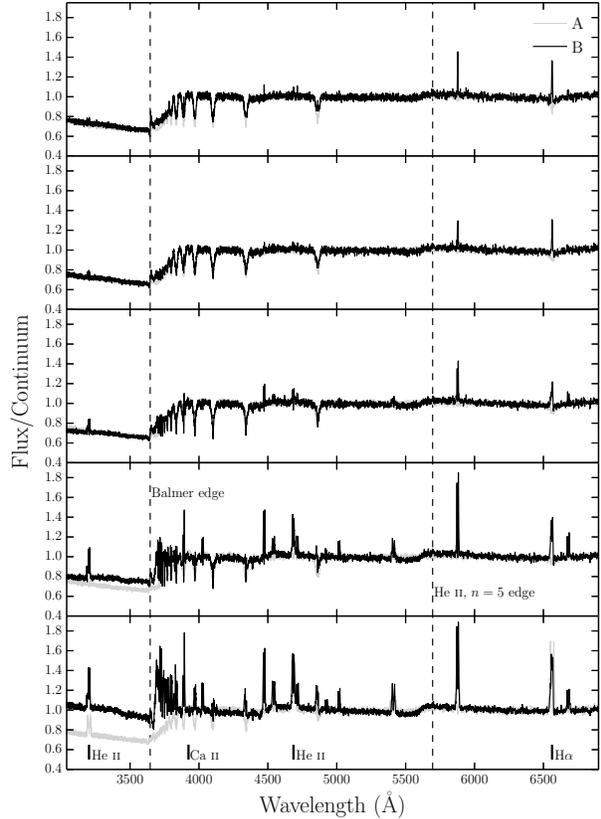}
\caption{
Synthetic optical spectra from model B computed for 
sightlines of 10, 27.5, 45, 62.5 and 80 degrees. 
Model A is shown in grey for comparison.
In these plots the flux is divided by a polynomial fit to the 
underlying continuum redward of the Balmer edge, so that 
line-to-continuum ratios and the true depth of the
Balmer jump can be shown.
}
\label{continuumb}
\end{figure}

\begin{figure} 
\includegraphics[width=0.45\textwidth]{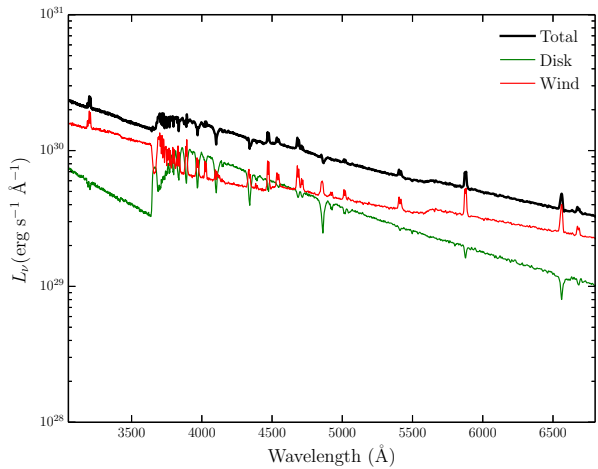}
\caption{Total packet-binned spectra across all viewing angles, in units
of monochromatic luminosity. 
The thick black line shows the total 
integrated escaping spectrum, 
while the green line shows disk photons which escape without being reprocessed by
the wind. The red line show the contributions from reprocessed 
photons. 
In this denser model the reprocessed contribution is significant compared
to the escaping disk spectrum. The Balmer continuum emission is prominent, and
the wind has a clear effect on the overall spectral shape.}
\label{modelb_escape}
\end{figure}

The wind now also has a clear effect on the continuum shape,
as shown by Fig.~\ref{modelb_escape}. In fact, the majority of the
escaping spectrum has been reprocessed in some way by the wind,
either by electron scattering (the wind is now moderately Thomson-thick),
or by bound-free processes. This is demonstrated by the flatter spectral shape
and the slight He photoabsorption edge present in the optical spectrum 
(marked in Fig.~\ref{continuumb}). This reprocessing is also
responsible for the change in continuum level between models A and B.
In addition, Figures~\ref{uvoptb}, \ref{continuumb} 
and \ref{modelb_escape} clearly demonstrate that the wind produces
a recombination continuum sufficient to completely fill in the Balmer jump
at high inclinations.\footnote{Note that the apparent absorption feature 
just redward of the Balmer jump in these models is artificial. It is
caused by residual line blanketing in the stellar atmospheres, which
our models cannot fill in since they employ a 20-level H atom.}
This might suggest that Balmer continuum emission from a wind can be important 
in shaping the Balmer jump region, as
originally suggested by Knigge et al.
(1998b; see also Hassall et al. 1985)\nocite{KLWB98,hassall}. 

It should be acknowledged, however,
that the Balmer jump in high-state CVs would naturally weaken at
high inclinations due to limb darkening effects \citep{ladous1989, ladous1989b}. 
Although we include a simple limb darkening law which affects 
the emergent flux at each inclination, we do not
include it as a {\em frequency dependent} opacity in our model.
As a result, the efficiency of filling in the Balmer jump
should really be judged at low and medium inclinations, 
where, although prominent, the recombination continuum does
not overcome the disk atmosphere absorption. 
In addition, this effect 
could mean that any model which successfully fills in the 
jump at low inclinations could lead to a Balmer jump 
in emission at high inclinations.
In any case, to properly understand this phenomenon, a fully self-consistent
radiative transfer calculation of both the disk atmosphere
and connected wind is required.

\begin{figure}
\includegraphics[width=0.5\textwidth]{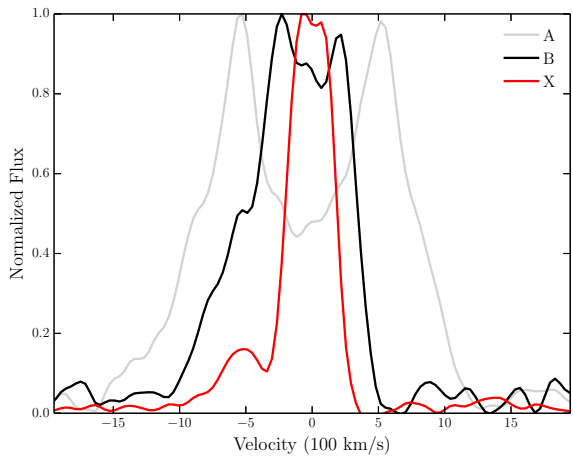}
\caption{
\ha\ line profiles, normalized to 1, plotted in velocity space 
for three models with varying kinematic 
properties, computed at an inclination of $80^\circ$.
The benchmark model and the improved optical
model described in section 6 are labeled as A and B respectively,
and a third model (X) which has an increased acceleration length of 
$R_v = 283.8~R_{WD}$, and $\alpha=4$ is also shown. 
The $x$-axis limits correspond to the Keplerian velocity at 
$4R_{WD}$, the inner edge of the wind.
We observe a narrowing of the lines, and a single-peaked line in model X.
This is not due to radial velocity shear (see section 5.3).
}
\label{halpha}
\end{figure} 

\subsection{Line Profile Shapes: Producing Single-Peaked Emission}

Fig.~\ref{halpha} shows how the H$\alpha$ profile changes with the kinematics of the wind for 
an inclination of $80^\circ$. The main prediction is that dense, slowly accelerating 
wind models produce narrower emission lines. This is {\em not} due to radial 
velocity shear. As stated by MC96, that mechanism can only work if poloidal 
and rotational velocity gradients satisfy $(dv_l/dr)/(dv_\phi/dr) \gtrsim 1$; in 
our models, this ratio is always $\lesssim 0.1$. Instead, the narrow lines predicted 
by our denser wind models can be traced to the base of the outflow becoming optically 
thick in the continuum, such that the line emission from the base of the wind
cannot escape to the observer. In such models, the `line photosphere'
(the $\tau \simeq 1$ surface of the line-forming region) moves outwards, towards larger 
vertical and cylindrical distances. This reduces the predicted line widths, since the 
rotational velocities -- which normally provide the main line broadening mechanism at 
high inclination -- drop off as $1/r$. This is not to say that the MC96 
mechanism could not be at work in CV winds. For example, it would be worth investigating
alternative prescriptions for the wind velocity field, as well as the possibility that the 
outflows may be clumped. An inhomogeneous flow 
(which has been predicted in CVs; see section 5.2)
might allow large radial velocity shears to exist while still 
maintaining the high densities needed to produce the required level of emission.
However, such an investigation is beyond the scope of the present paper.

In our models, single-peaked line profiles are produced once the line forming region has been
pushed up to $\sim 10^{11}$~cm ($\sim150~R_{WD}$) above the disk plane. 
This number may seem unrealistically large, but the vertical extent of 
the emission region is actually not well constrained observationally. 
In fact, multiple observations of eclipsing NLs show that the H$\alpha$ 
line is only moderately eclipsed compared to the continuum (e.g. Baptista et al. 2000;
Groot et al. 2004; see also section 5.4), 
implying a significant vertical extent for the line-forming 
region. This type of model should therefore not be ruled out {\em a priori}, 
but this specific model was not adopted as our optically optimized model
due to its unrealistically high continuum level in eclipse. 

\subsection{Sensitivity to Parameters}

This revised model demonstrates that one can achieve a more
realistic optical spectrum by altering just two kinematic parameters. 
However, it may also be possible to achieve this by modifying
other free parameters such as $\dot{M}_{wind}$, the opening angles of the wind and the 
inner and outer launch radii. For example, increasing the mass loss rate of the wind
increases the amount of recombination emission (which scales as $\rho^2$), 
as well as lowering the ionization parameter and increasing optical depth through the wind. 
Wider launch radii and opening angles lead to a larger emitting volume, but this is moderated by a decrease in density 
for a fixed mass-loss rate. We also note that the inner radius of $4~R_{WD}$ adopted by SV93 
affects the emergent UV spectrum seen at inclinations $<\theta_{min}$ as 
the inner disk is uncovered. This causes less absorption in the UV resonance lines,
but the effect on the optical spectrum is negligible.
We have verified this general behaviour, but
we suggest that future work should investigate the effect of these parameters in more detail
as well as incorporating a treatment of clumping.
If a wind really does produce the line and continuum emission seen in optical spectra of high-state CVs, then
understanding the true mass loss rate and geometry of the outflow is clearly important.

\subsection{Comparison to RW Tri}

\begin{figure*}
\includegraphics[width=0.9\textwidth]{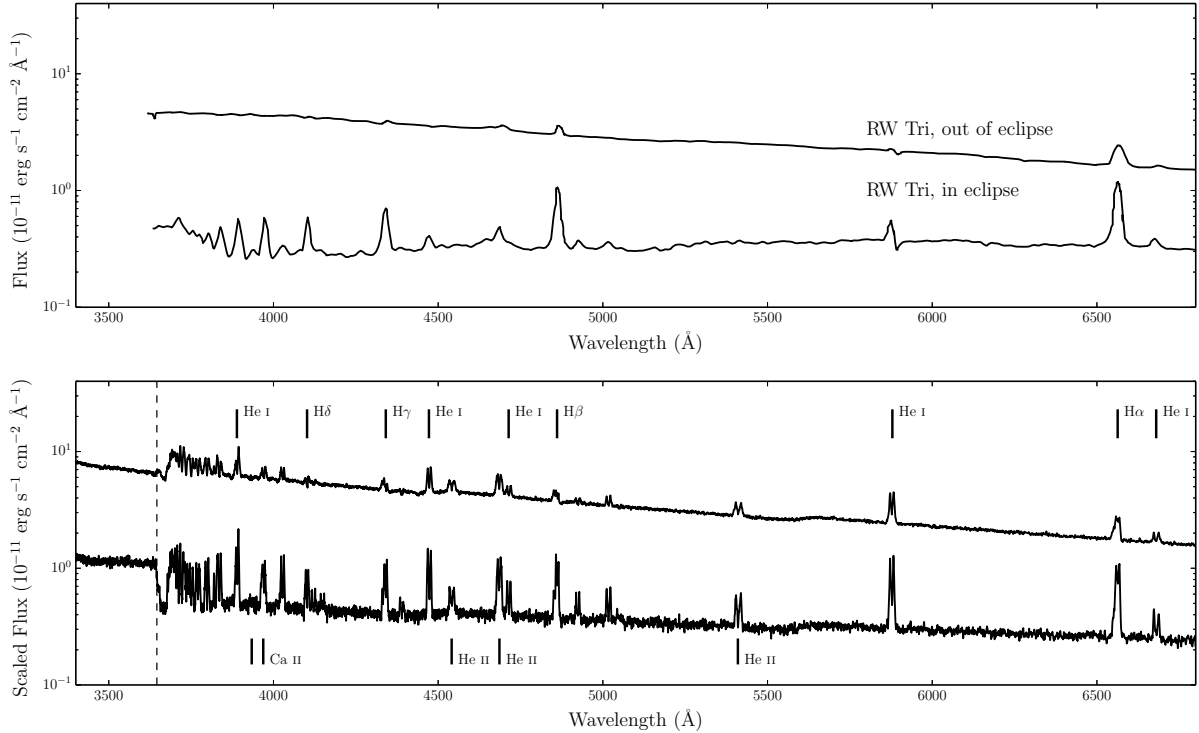}
\caption{{\sl Top Panel:} In and out of eclipse spectra of the high
inclination NL RW Tri. {\sl Bottom Panel:} In and out of eclipse synthetic
spectra from model B.
The artificial `absorption' feature just redward of the Balmer jump
is caused for the reasons described in section 5.2.}
\label{rwtricomp}
\end{figure*}

Fig.~\ref{rwtricomp} shows a comparison of the predicted
out-of-eclipse and mid-eclipse spectra against observations of the
high-inclination nova-like RW~Tri. The inclination of RW Tri is
somewhat uncertain, with estimates including $70.5^\circ$
\citep{smak1995}, $75^\circ$ \citep{groot2004}, $80^\circ$
\citep{longmore1981} and $82^\circ$\citep{frankking1981}. Here, we
adopt $i = 80^\circ$, but our qualitative conclusions are not
particularly sensitive to this choice. 
We follow LK02 is setting the value of $r_{disk}$ (the maximum radius of the accretion disk)
to $34.3~R_{WD}$. When compared to the semi-major axis of RW Tri,
this value is perhaps lower than one might 
typically expect for NLs \citep{harropallinwarner1996}. 
However, it is consistent
with values inferred by \cite{rutten1992}.
We emphasize that this model is in no sense a fit to this -- or any other -- data set.

The similarity between the synthetic and observed spectra is
striking. In particular, the revised model produces strong emission in
all the Balmer lines, with line-to-continuum ratios comparable to
those seen in RW Tri. Moreover, the line-to-continuum contrast
increases during eclipse, as expected for emission produced in a disk
wind. This trend is in line with the observations of RW~Tri, and it
has also been seen in other NLs, including members of the SW~Sex class
\citep{neustroev2011}. As noted in section 5.2, the majority
of the escaping radiation has been reprocessed by the wind in some way
(particularly the eclipsed light).

However, there are also interesting differences between the revised
model and the RW Tri data set. For example, the model exhibits
considerably stronger He~{\sc ii} features than the observations,
which suggests that the overall ionization state of the model is
somewhat too high. As discussed in section 5.3, the optical lines are
narrow, but double-peaked. This is in contrast to what is generally seen in observations
of NLs, although the relatively low resolution of the RW Tri
spectrum makes a specific comparison difficult. In order to demonstrate
the double-peaked nature of the narrower lines, we choose not to 
smooth the synthesized data to the resolution of the RW Tri dataset.
If the data was smoothed, the \ha\ line would appear single-peaked.

%
%

\section{Conclusions}

We have investigated whether a disk wind model designed to reproduce
the UV spectra of high-state CVs would also have a significant effect
on the optical spectra of these systems. We find that this is indeed
the case. In particular, the model wind produces H and He
recombination lines, as well as a recombination continuum blueward of
the Balmer edge. We do not produce P-Cygni profiles
in the optical H and He lines, 
which are seen in a small fraction of CV optical spectra.
Possible reasons for this are briefly discussed in section 
5.2.

We have also constructed a revised benchmark model which is designed
to more closely match the optical spectra of high-state CVs. This
optically optimized model produces all the prominent optical lines in
and out of eclipse, and achieves reasonable verisimilitude with the
observed optical spectra of RW Tri. However, this model also has
significant shortcomings. In particular, it predicts
stronger-than-observed He~{\sc ii} lines in the optical region and too
much of a collisionally excited contribution to the UV resonance lines. 

Based on this, we argue that recombination emission 
from outflows with sufficiently high densities and/or optical depths 
might produce the optical lines observed in CVs, and may also 
fill in the Balmer absorption edge in the spectrum of the accretion disk, 
thus accounting for the absence of a strong edge in observed CV spectra.
In section 5.3, we demonstrate that
although the double peaked lines narrow and 
single-peaked emission can be formed in our densest models, 
this is not due to the radial velocity shear mechanism proposed by MC96.
We suggest that `clumpy' line-driven winds or a different
wind parameterization may nevertheless allow this mechanism to work.
We also note the possibility that, as in our denser models, 
the single-peaked lines are formed well above the disk, where 
rotational velocities are lower.

It is not yet clear whether a wind model such as this can
explain all of the observed optical features of high-state CVs --
further effort is required on both the observational
and modelling fronts.
However, our work demonstrates that {\sl disk winds matter}. They are
not just responsible for creating the blue-shifted absorption and
P-Cygni profiles seen in the UV resonance lines of high-state CVs, but
can also have a strong effect on the optical appearance of these
systems. In fact, most of the optical features characteristic of CVs
are likely to be affected -- and possibly even dominated -- by their disk
winds. Given that optical spectroscopy plays the central role in
observational studies of CVs, it is critical to know 
where and how these spectra are actually formed. We believe it is high
time for a renewed effort to understand the formation of spectra in
accretion disks and associated outflows.

\subsection*{Acknowledgements}
The work of JHM and CK is supported by the Science and Technology Facilities Council (STFC), 
via studentships and a consolidated grant, respectively. 
The work of NSH is supported by NASA under Astrophysics Theory Program grants NNX11AI96G  and NNX14AK44G.
We would like to thank the anonymous referee for a helpful and constructive report, and 
we are  grateful to A.F. Palah and B.T. Gaensicke for the IX Vel XSHOOTER dataset.
We would also like to thank J.V. Hernandez Santisteban, S.W. Mangham and I. Hubeny for useful discussions. 
We acknowledge the use of the IRIDIS High Performance Computing Facility, 
and associated support services at the University of Southampton, in the completion of this work.

\bibliography{mybib.bib}

\end{document}